\documentclass[12pt]{iopart}
\usepackage{a4, exscale, latexsym, amssymb, isolatin1, fancyheadings} 
\usepackage[latin1]{inputenc}
\usepackage{amssymb}
\usepackage{epsf}
\usepackage{epsfig}

\newcommand{\sts}{\scriptsize}
\newcommand{\mb}{\mbox}

\begin{document}

\title[Microcanonical specific heat]{Finite-size behaviour of the microcanonical
specific heat}

\author[H Behringer \etal]{H Behringer, M Pleimling and A H\"{u}ller}

\address{Institut f\"ur Theoretische Physik I, Universit\"at
Erlangen-N\"urnberg, D-91058 Erlangen, Germany}

\ead{hans.behringer@physik.uni-erlangen.de}

\begin{abstract}
For models which exhibit a continuous 
phase transition in the thermodynamic limit a numerical study of small systems reveals a 
non-monotonic  behaviour of the microcanonical specific heat 
as a function of the system size. This is
in contrast to a treatment in the canonical ensemble where the maximum 
of the specific heat increases monotonically with the size of the system.
A phenomenological theory is developed which permits to describe this
peculiar behaviour of the microcanonical specific heat and allows in
principle the determination of microcanonical critical exponents.
\end{abstract}


\section{Introduction}
In recent years numerous studies investigated the possible differences
between the microcanonical and the canonical treatment of a given system.
It is now well accepted that in various cases the
microcanonical and the canonical ensemble are not equivalent
\cite{thirr70,Hue94a,elli00,daux00,Gro01b,Bar01,Isp01,gulm02}. 
For short-range interactions,
the equivalence of the two ensembles holds in the infinite volume limit,
but this is not the case in finite systems. For long-range interactions,
as encountered for example in gravitational systems, the two ensembles
remain inequivalent even for infinite systems.

Clearly, this inequivalence in finite systems makes the microcanonical analysis of possible
signatures of phase transitions an important issue \cite{Hue94a,kast00,Gro00b,Gro01b,huel02,plei04}. 
For discontinuous phase transitions,
the microcanonical analysis reveals typical small system signatures, as
e.\,g. a back-bending
of the caloric curve or the appearance of a negative specific heat.  
Negative heat capacities have indeed been measured in recent
experiments on nuclear fragmentation \cite{Dag00} and on the
melting of atomic clusters \cite{Sch01a}. Similarly, intriguing features are also revealed
in the microcanonical analysis of small systems which 
exhibit a continuous phase transition in the thermodynamic limit. Indeed, typical features of symmetry breaking,
as e.\,g. the abrupt onset of a non-zero order parameter
when the (pseudo-)critical point is approached from above or a diverging
susceptibility, turn up
already for finite systems \cite{kast00}. This is in contrast
to the canonical ensemble where singularities
appear exclusively in the thermodynamic limit. 

The fundamental quantity in a microcanonical analysis is the density of states or,
equivalently, the microcanonical entropy. All relevant quantities can indeed be
expressed by partial derivatives of the microcanonical entropy. For example,
the susceptibility is proportional to the inverse of the curvature of the entropy surface. It is
the existence of a point with vanishing curvature that is responsible
for the divergent susceptibility observed in finite systems which have a continuous phase transition
in the infinite volume limit.

In the present work we examine more closely the finite-size behaviour of the microcanonical
specific heat in different classical spin systems. For systems with a continuous phase
transition one expects that the maximum of the specific heat increases with increasing system
size. This is indeed observed in the microcanonical analysis for not too small systems.
For small system sizes, however, we observe a non-monotonic behaviour as the
maximum of the specific
head first {\it decreases} for increasing system sizes. This is again a property of the
entropy surface as the microcanonical specific heat can be exclusively expressed by
energy derivatives of the microcanonical entropy. In order to account for this
peculiar behaviour we develop a phenomenological theory based on the analyticity of
the entropy surface of finite systems.

The paper is organized as follows. In Section 2 we first discuss the
definition of the temperature in the microcanonical ensemble.
In the microcanonical ensemble various definitions of the temperature 
are possible, the different expressions becoming equivalent in the thermodynamic limit.
The microcanonical specific heat, based on the expressions for the temperature,
is the subject of Section 3. Numerical results obtained for two- and three-dimensional 
Ising models as well as for the two-dimensional three-state Potts model reveal
a non-monotonic behaviour of the specific heat for increasing system sizes.
The finite-size behaviour of the specific heat of 
microcanoncial systems is considered from a phenomenological point of view in
Section 4 where a finite-size scaling theory is developed
which explains the peculiar behaviour of the microcanonical specific heat of small
systems. Finally, Section 5 gives our conclusions.

\section{Temperature in the microcanonical ensemble}
The density of states is the starting point for the statistical
description of thermostatic properties in the different ensembles.
For a magnetic system that is isolated from any environment the proper
natural variables are the energy $E$ and the magnetisation $M$.
The corresponding characteristic function of the isolated system is the
microcanonical entropy
\begin{equation}
    \label{eq:def_entropie}
    S(E, M, L^{-1}) = \ln \Omega(E,M, L^{-1}) 
\end{equation}
where $\Omega$ denotes the degeneracy of the macrostate $(E,M)$ and $L$
is the linear extention of the system.
Here and in the following units with $k_B = 1$ are used.
The microcanonical analysis of finite classical spin systems
starts from the microcanonical entropy density
\begin{equation}
    s(e,m, L^{-1}) := \frac{1}{L^d} S(L^de, L^dm, L^{-1})
\end{equation}
of a system in $d$ dimensions with $N=L^d$ spins, where $e=E/N$ denotes the energy density and
$m=M/N$ the magnetisation density. In the following
the dependence on the system size is suppressed in order to improve
readability. 

Before investigating the microcanonical specific heat, we first have to discuss
the definition of the temperature for finite systems in
the microcanonical ensemble. 
In the thermodynamic limit canonically defined physical quantities and the
corresponding microcanonical quantities have to become identical for systems
with suitably short range forces. However,
this requirement does not yield an unambiguous definition of the
microcanonical temperature, leading to different physically plausible definitions 
in finite systems which all become equivalent in the thermodynamic limit. 

The starting point is the canonical partition
function
\begin{equation}
    \label{eq:spez_zustandssumme}
    Z(\tilde{\beta},\tilde{h}) = L^{2d}\int \mb{d}e\,  \int \mb{d}m \, \exp\left\{
    L^d ( s(e,m) -  \tilde{\beta} e +  \tilde{\beta} \tilde{h}m)\right\}
\end{equation}
which is the Laplace transform of the density of states. Here the inverse
canonical temperature $\tilde{\beta}$ and the applied magnetic
field  $\tilde{h}$ are external
parameters which are imposed on the system by its environment. The canonical
temperature and external field are denoted by a tilde in order to avoid
any confusion with the microcanonical temperature and field defined in the
following. The integral (\ref{eq:spez_zustandssumme}) can be
evaluated in the limit $L\to \infty $ by means of the Laplace
method. For a given inverse temperature $\tilde{\beta}$ and
external magnetic field $\tilde{h}$ the dominant contributions to
the integral arise from the maximum of the argument $g(e,m) =
s(e,m) - \tilde{\beta} e +  \tilde{\beta}
 \tilde{h}m.$
The equations $\partial_eg=0$ and $\partial_mg=0$ suggest the following
definitions of the inverse microcanonical temperature:
\begin{equation} \label{beta1}
\beta(e,m) = \partial_e s(e,m)
\end{equation}
and of the microcanonical magnetic field:
\begin{equation}
        \beta(e,m) h(e,m) = - \partial_m s(e,m).
\end{equation}
Here and in the following the notation $\partial_x$ is used for the partial derivative
$\partial/\partial x$. 
The inverse microcanonical temperature $\beta$ and magnetic field $h$
are conjugate variables of the natural variables $e$ and $m$ of
the microcanonical approach and consequently depend on these. The
definition (\ref{beta1}) of the microcanonical temperature surface leads to
the follwing definition of the temperature in equilibrium. Consider
the spontaneous magnetisation $m_{\sts \mb{sp}}(e)$ of the
magnetic system for a given energy $e$ that is defined by the condition
$h(e,m) = 0$ \cite{kast00}. 
The temperature of the magnetic system in equilibrium is then obtained by
evaluating the inverse temperature surface $\beta(e,m)$ at the equilibrium
macrostate $(e,m_{\sts \mb{sp}}(e))$:
\begin{equation}
\label{gl:gleichtemp}
\beta_{\sts \mb{E}}(e) := \beta(e,m_{\sts \mb{sp}}(e)) = \partial_e
s(e,m)|_{m=m_{\sts \mb{sp}}(e)}.
\end{equation}
This definition of the inverse temperature ensures ensemble equivalence between the canonical and
microcanonical description, as can be seen using the Laplace method in the
asymptotic limit $L\to\infty$. For finite $L$, however, the exponential in
(\ref{eq:spez_zustandssumme}) cannot be approximated by the
quadratic term of the Taylor expansion only. Higher order terms are necessary
which render the integrand asymmetric. Consequently, the canonical mean
values are shifted from the associated maximum of the entropy surface
leading to the inequivalence of the canonical and the microcanoncial
ensemble for finite system sizes.   

We pause here for a moment to recall that
the spontaneous magnetisation  $m_{\sts \mb{sp}}(e)$ of a finite
microcanonical magnetic system exhibits features which are typical of phase
transitions. The spontaneous magnetisation 
of the Ising model in dimensions $d \geq 2$, for example,  is
zero above a well-defined transition energy $e_{\sts \mb{pc}}$ and
becomes non-zero below $e_{\sts \mb{pc}}$. Close to this pseudo-critical
energy the variation of the spontaneous magnetisation as a
function of the deviation of $e$ from $e_{\sts \mb{pc}}$ is described by a square root function~\cite{kast00,huel02}.
This classical behaviour has its origin in the
analyticity of the entropy surface for all finite systems~\cite{behr04,gros04}.
The appearance of a non-zero spontaneous magnetisation
reflects the spontaneous breakdown of the global symmetry of the
system and may be regarded as a precursor
of the critical point of the infinite system~\cite{kast00,huel02,behr03}.
Note that the specific
entropy $s_\infty(e,m)$ in the thermodynamic limit is a concave function of
its variables. In finite systems, however, this is not compulsory so that
two maxima of the entropy  can appear at non-zero magnetisations for a given energy.

Coming back to the canonical ensemble we remark that in absence of a magnetic
field the partition function (\ref{eq:spez_zustandssumme}) simplifies to
\begin{equation}
    \label{eq:spez_zustandssumme2}
    Z(\tilde{\beta},\tilde{h}=0) = L^{d}\int \mb{d}e\,
\left(L^d \int \mb{d}m \, \exp  \left\{
    L^d  s(e,m)\right\}\right)   \exp \{ -L^d\tilde{\beta} e\},
\end{equation}
which leads to  the definition
\begin{equation}
\label{gl:definiereSR}
    \exp\{ L^d s_{\sts \mb{R}}(e)\} = L^d \int \mb{d}m \, \exp  \left\{
    L^d  s(e,m)\right\}
\end{equation}
of the reduced (specific) entropy $s_{\sts \mb{R}}(e)$. In the limit of
large system sizes the
dominant contributions to the integral
(\ref{eq:spez_zustandssumme2}) arise from the energy defined by
the maximum of the argument $s_{\sts \mb{R}}(e) - \tilde{\beta}e$
for a given inverse canonical temperature $\tilde{\beta}$. This
suggests the following alternative definition of an inverse (reduced) microcanonical temperature,
namely
\begin{equation}
    \beta_{\sts \mb{R}}(e) = \frac{\mb{d}}{\mb{d}e} s_{\sts \mb{R}}(e).
\end{equation}
The thermal properties of the system are now obtained from the entropy
function  $ s_{\sts \mb{R}}(e)$ rather than from the full entropy surface $s(e,m)$
depending on both the energy and the magnetisation.

To conclude this section the interrelation between the inverse temperatures
$\beta_{\sts \mb{E}}(e)$ and $\beta_{\sts \mb{R}}(e)$ is briefly considered.
In the asymptotic limit $L\to \infty $ the integral
(\ref{gl:definiereSR}) is dominated by the entropy
$s(e,m_{\sts \mb{sp}}(e))$, evaluated at the spontaneous magnetization $m_{\sts \mb{sp}}(e)$,
as can again be seen by using the
Laplace method. 
Therefore, the entropy $ s_{\sts \mb{R}}(e)$ is given by $s(e,m_{\sts \mb{sp}}(e))$ for
asymptotically large system sizes $L$ and one gets
\begin{equation}
\label{gl:deftempasy}
\beta_{\sts \mb{R}}(e) \stackrel{L\to \infty}{\sim}\frac{\mb{d}}{\mb{d}e} s(e,m_{\sts \mb{sp}}(e)).
\end{equation}
Carrying out this
differentiation we obtain the relation 
\begin{equation}
\beta_{\sts \mb{R}}(e) \stackrel{L\to \infty}{\sim}
\frac{{\partial}}{{\partial}e} s(e,m) |_{m=m_{\sts \mb{sp}}(e)}  +
\frac{{\partial}}{{\partial}m} s(e,m) |_{m=m_{\sts \mb{sp}}(e)}
\frac{\mb{d}}{\mb{d}e} m_{\sts \mb{sp}}(e). 
\end{equation}
As $\partial_m s(e,m)$ is zero at $(e,m_{\sts \mb{sp}}(e))$ the second term vanishes
and one is left with $\beta_{\sts \mb{R}}(e) \sim \beta_{\sts \mb{E}}(e)$.
The full entropy surface $s(e,m)$ and the reduced entropy function
$s_{\sts \mb{R}}(e)$ will therefore lead to the same equilibrium temperature in
the asymptotic limit $L \to \infty$. For finite $L$, however, 
$\beta_{\sts \mb{E}}(e)$ is significantly different from 
$\beta_{\sts \mb{R}}(e)$. 

\section{Microcanonical specific heat}

\subsection{General discussion}

Once the inverse temperature $\beta$  of a microcanonical system is evaluated --- here $\beta$ may be $\beta_{\sts
\mb{E}}$ or  $\beta_{\sts
\mb{R}}$ --- one can
calculate the specific heat which
is generally defined by  $c=\mb{d}u/\mb{d}T$ with $u$
and $T = 1/\beta$ being the energy and the temperature of the system.
For the microcanonical specific heat as a function of the energy of the
system this gives 
\begin{equation}
\label{eq:spezallgfuermikro}
c(e) = -(\beta(e))^2\left(\frac{\mb{d}\beta(e)}{\mb{d}e}  \right)^{-1}
= -\left(\frac{\mb{d}s}{\mb{d}e}  \right)^2 \left(\frac{\mb{d}^2s}{\mb{d}e^2}  \right)^{-1}.
\end{equation}
The discussed ambiguity in the definition of the microcanonical temperature leads also
to different expressions for the specific heat in finite systems. However,
they converge towards the same limit function in the thermodynamic limit.

In the following we discuss the finite-size behaviour of 
the specific heat arising from the temperature
$\beta_{\sts \mb{R}}(e)$ in different classical spin models 
(from now on we drop the subscript R in order to avoid unnecessary
notation). Note that this is the definition of the specific heat that is the
most relevant for experiments where usually the energy is considered
as the unique natural variable corresponding to systems to which no external
field is applied. Specifically, we study three models undergoing a continuous
phase transition in the thermodynamic limt: the two- and  and the three-dimensional Ising model
as well as the three-state Potts model in two dimensions.
The nearest neighbour Ising model is defined by the Hamiltonian 
\begin{equation}
        \mathcal{H} = -\sum_{\left< i,j \right>}\sigma_i \sigma_j,
\end{equation}
where the summation over nearest neighbour
pairs is indicated by
$\left< i,j \right>$ and the spin $\sigma_i$ at site $i$ can be in  the states $\sigma_i = \pm
1$. In the present study Ising models defined on the square and on the cubic lattices are
considered.
The three-state Potts model is a generalisation of the Ising model where the Potts spins $\sigma_i$
take on the values $1, 2, 3$.
The Hamiltonian is given by
\begin{equation}
        \label{eq:potts_hamilton}
	  \mathcal{H} = -\sum_{\left< i,j \right>} \delta_{\sigma_i,
\sigma_j} ,
\end{equation}
where $\delta_{\sigma_i,\sigma_j} = 1$ when the spins located at the neighbouring sites
$i$ and $j$ have the same value and zero otherwise. For the Potts model we
only consider the square lattice.

In finite systems the appearance of a continuous phase
transition in the thermodynamic limit is signaled by a maximum in the
specific heat which becomes more and more pronounced when the system size is
increased. This behaviour of the specific heat is due to a maximum of the
second derivative of $s(e)$ which is negative everywhere and tends to zero
for increasing system sizes from below. The position of the maximum of
the second derivative of $s(e)$ defines a pseudo-critical energy $e_{\sts
\mb{pc}}$ of the finite system. At the same time the microcanonical inverse
temperature $\beta_{\sts \mb{pc}}:=\beta(e_{\sts \mb{pc}})$ evaluated at the
energy $e_{\sts \mb{pc}}$ converges towards the critical value $1/T_{\sts
\mb{c}}$ 
when $L$ tends to infinity.

\begin{figure}[h!]
\begin{center}
\epsfig{file=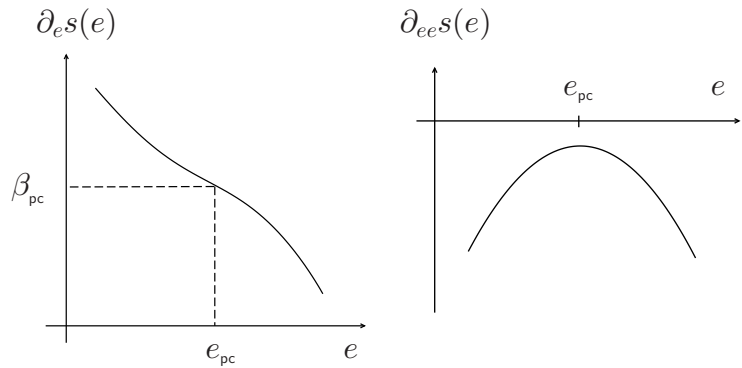,width=13em, angle=270}
\caption[Schematic depiction of the derivatives of the reduced
entropy]{\label{bild:schema_redtempspez}\small Schematic depiction
of the first two derivatives of the reduced entropy near the pseudo-critical
energy of a finite system. The pseudo-critical energy $e_{\sts \mb{pc}}$ corresponds
to an inverse temperature $\beta_{\sts \mb{pc}}$
(left). The derivative $\partial_{ee}s$ has a maximum at $e_{\sts
\mb{pc}}$ (right). }
\end{center}
\end{figure}

The behaviour just described, which is schematically sketched in figure
\ref{bild:schema_redtempspez}, is indeed observed in the different models for
not too small system sizes. For very small systems, however, our numerical
results reveal an unexpected non-monotonic behaviour of the specific heat,
as discussed in the next subsection.

\subsection{Numerical results}
\label{kap:untersuchespezskal}

\begin{figure}
\begin{center} 
\epsfig{file=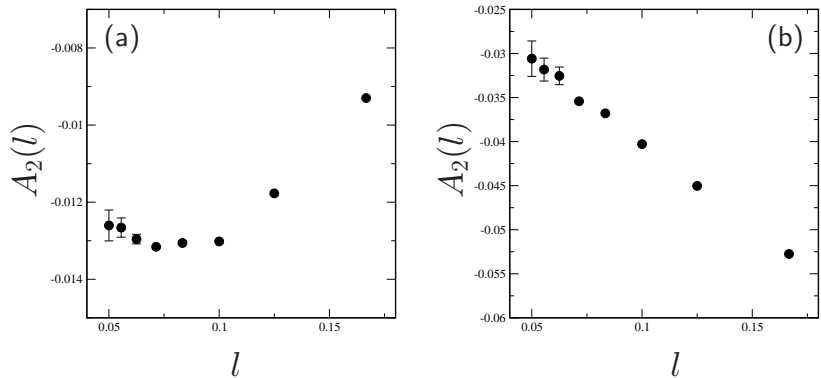,width=13em, angle=270}
\caption[Evolution of the coefficient $A_2(l)$ for finite
three-dimensional Ising systems
]{\label{bild:is3d_koeffevolution}\small Evolution of the
coefficient $A_2(l)$ for finite three-dimensional Ising systems
with periodic boundary conditions (a) and open boundaries (b). The
modulus of the coefficient $A_2(l)$ of the system with periodic
boundary conditions has a plateau and becomes smaller again for
small systems. In the thermodynamic limit $l\to 0$ both curves have to
extrapolate to $A_2=0$.}
\end{center}
\end{figure}

In this subsection the specific heat of finite Ising and Potts systems is
investigated numerically for both periodic and open boundary conditions. 
To obtain the numerical data we used a recently proposed very efficient method for the 
direct computation of the density of states~\cite{huel02}.
Specifically, we discuss in the following
the value $\partial_{ee}s(e_{\sts
\mb{pc}},L^{-1})$ of the second derivative of the entropy evaluated at the
pseudo-critical energy $e_{\sts \mb{pc}}$. For later
convenience this value is denoted by $A_2(l)$ where $l:=L^{-1}$ is the inverse
system size:
\begin{equation} \label{a2l}
 A_2(l) = \partial_{ee}s(e_{\sts
\mb{pc}},l).
\end{equation}

The coefficient $A_2(l)$
of finite three-dimensional Ising systems is shown in
figure \ref{bild:is3d_koeffevolution}. With
periodic boundary conditions the coefficient $A_2(l)$ shows a
back-bending as its modulus first increases for increasing $l$ (i.\,e.\ decreasing system sizes)
and then decreases for very small systems, see figure~\ref{bild:is3d_koeffevolution}a. This
intriguing and unexpected
back-bending is not observed in the system with open boundaries.
 
\begin{figure}[h!]
\begin{center} \epsfig{file=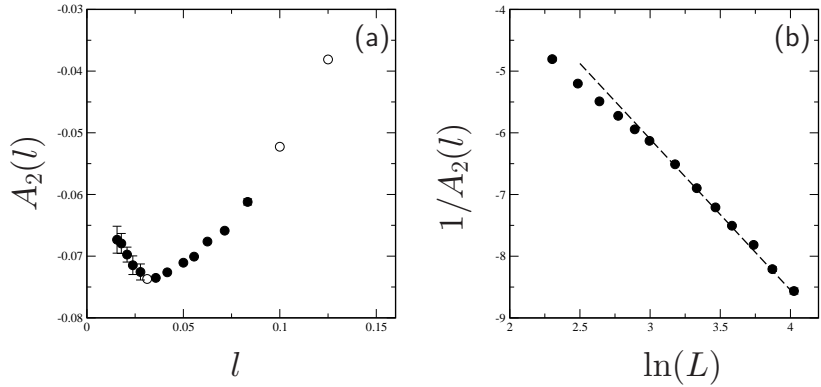,width=13em,
angle=270}
\caption[Evolution of the coefficient $A_2(l)$ for finite Ising
models in two dimensions] {\label{bild:is2d_koeffevolution}\small
Evolution of the coefficient $A_2(l)$ for finite Ising models in
two dimensions with periodic boundary conditions (a) and free
boundaries (b). The open circles in (a) display exactly evaluated data. 
The coefficient $A_2(l)$ bends back for small
systems only for periodic boundary conditions. For the system with
open boundaries a logarithmic plot shows that the coefficient
$A_2(l)$ evolves logarithmically for large systems (see also section 4).}
\end{center}
\end{figure}
 
Similarly, the coefficient $A_2(l)$ of the two-dimensional Ising model with
periodic boundary conditions also exhibits this back-bending for very
small system sizes, whereas again no back-bending is observed for
open boundaries, see figure~\ref{bild:is2d_koeffevolution}. In case of the
systems with linear extensions $L=8,10$ and $32$ and periodic boundaries the numericaly
determined data can be compared to exactly computed data
\cite{bind72,cres95,beale96}. This is also
indicated in figure \ref{bild:is2d_koeffevolution}.  

\begin{figure}[h!]
\begin{center}
\epsfig{file=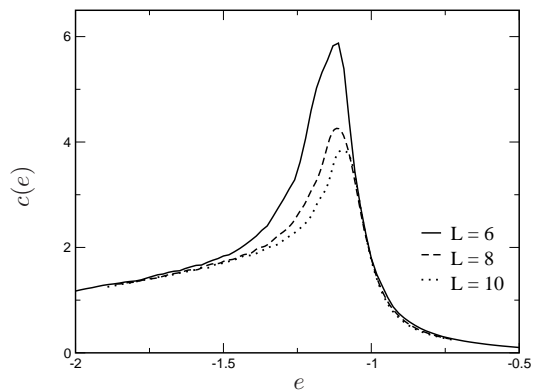,width=13em, angle=270}
\caption[Evolution of the microcanonical specific heat for small
three-dimen\-sional Ising systems
]{\label{bild:3dis_spezevolution}\small Evolution of the
microcanonical specific  heat for small three-dimensional Ising
systems with periodic boundary conditions. The maximum decreases
for increasing system size. }
\end{center}
\end{figure}

Naturally, the back-bending of the coefficient $A_2(l)$ directly affects the
behaviour of the microcanonical specific heat of small systems as can be
seen from equations
(\ref{a2l}) and (\ref{eq:spezallgfuermikro}). Indeed,
the maximum of the specific heat first
{\it decreases} with growing system size before increasing again,
thus yielding a divergence in the
thermodynamic limit. This decrease of the specific heat of small
microcanonical systems is displayed in figure
\ref{bild:3dis_spezevolution} for the three-dimensional Ising
model with periodic boundary conditions. It is worth noting that such a peculiar behaviour
of the specific heat of small systems is not observed in the
canonical ensemble (see, e.\,g.\,\cite{bind72,land76a,land76b}). 

\begin{figure}[h!]
\begin{center}
\epsfig{file=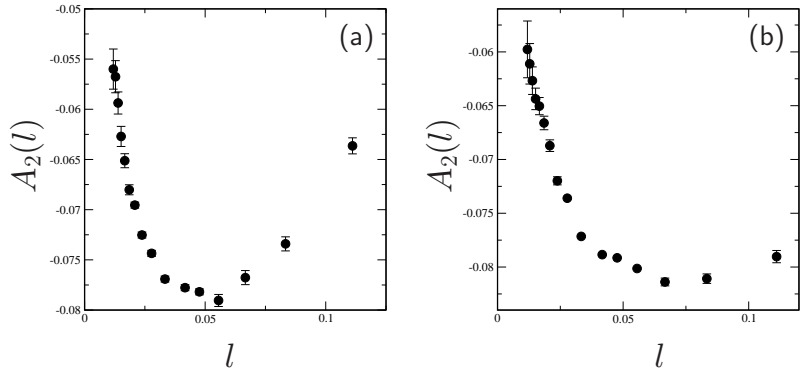,width=13em, angle=270}
\caption[Evolution of the coefficient $A_2(l)$ for finite
three-state Potts models in two dimensions
]{\label{bild:po2d_koeffevolution}\small Evolution of the
coefficient $A_2(l)$ for finite three-state Potts models in two
dimensions with periodic boundary conditions (a) and free
boundaries (b). For both boundary conditions the coefficient
$A_2(l)$ bends back for small systems. 
}
\end{center}
\end{figure}

Finally, figure \ref{bild:po2d_koeffevolution} displays the evolution of the
coefficient $A_2(l)$ for finite two-dimensional three-state Potts models for both 
periodic and open boundary conditions. The
back-bending is strongly pronounced for periodic
boundaries and is in this case also visible, but less developed, for
free boundaries. 
For large systems the coefficient $A_2(l)$ eventually approaches 
zero reflecting the appearance of a continuous transition in the
infinite system. 
 
\section{Phenomenological theory for finite systems}

In this section the behaviour of the specific heat of finite microcanonical
systems is investigated from a phenomenological point of view. As discussed in the following 
this leads to
a theoretical description that accounts for the peculiar behaviour of the microcanonical
specific heat described in the previous section. 

The specific heat of the infinite Ising or Potts systems diverges at the
critical point $e_{\sts \mb{c}, \infty}$. For a system with a power law singularity the specific heat
has the form 
\begin{equation}
c_\infty(e) \sim |e-e_{\sts \mb{c},\infty}|^{-\alpha_\varepsilon}
\end{equation}
in the vicinity of the critical point, where $\alpha_\varepsilon$ denotes
the microcanonical critical exponent. The specific entropy of the
infinite system contains a singular part that is a generalised homogeneous
function in the vicinity of $e_{\sts \mb{c},\infty}$ characterised by the degree of
homogeneity $a_\varepsilon$~\cite{kast00}. The microcanonical
critical exponent $\alpha_\varepsilon$ is related to $a_\varepsilon$ by 
\begin{equation}
\alpha_\varepsilon =
\frac{1-2a_\varepsilon}{a_\varepsilon} = \frac{\alpha}{1-\alpha} 
\end{equation}
with $\alpha$ being the critical exponent of the canonical specific heat~\cite{kast00}.
Similarly, the microcanonical critical exponent $\nu_\varepsilon$ of the
correlation length can be expressed as 
\begin{equation}
\nu_\varepsilon  = \frac{1}{d a_\varepsilon}= \frac{\nu}{1-\alpha}
\end{equation}
where the dimensionality of the system is again denoted by $d$.

The discussion of finite-size scaling relations of the specific heat starts from the
decomposition
\begin{equation}
    \label{eq:Mffs_zerlegung}
    s(\varepsilon,l) = s_{\sts \mb{r}}(\varepsilon,l) + s_{\sts \mb{s}}(\varepsilon,l) \;
\end{equation}
of the entropy of a finite system into a regular and a singular part. The
deviation of the energy from the pseudo-critical energy is denoted by
$\varepsilon:= e - e_{\sts \mb{pc}}$, where $l$  again is the inverse
system size. The singular and regular
parts of the entropy of the finite systems are chosen to approach
the corresponding singular and regular part 
of the entropy of the infinite lattice:
\begin{equation}
    \lim_{l\to 0}s_{\sts \mb{s/r}}(\varepsilon,l) =s_{\infty, \sts \mb{s/r}}(\varepsilon) .
\end{equation}
Note that the singular part of the entropy of a finite system is an analytic
function due to the analyticity of thermodynamic potentials of finite
systems. 
The singular part $s_{\sts \mb{s}}(\varepsilon, l)$ is
assumed to obey the scaling assumption
\begin{equation}
    \label{eq:Mffs_annahme}
        s_{\sts \mb{s}}(\varepsilon,l) = \frac{1}{\lambda} s_{\sts \mb{s}}(\lambda^{a_{\varepsilon}}\varepsilon,
        \lambda^{1/d}l)
\end{equation}
with a positive re-scaling factor $\lambda$ and the degree of
homogeneity $a_{\varepsilon}$ discussed above. The ansatz
(\ref{eq:Mffs_annahme}) for the finite-size behaviour of the
singular part of the microcanonical entropy of finite systems does
not account for additional finite-size corrections that arise from
the contributions of irrelevant scaling fields. The qualitative
picture that is developed in the following can be extended to
include those contributions as well. Differentiating the
finite-size scaling assumption (\ref{eq:Mffs_annahme}) with
respect to the re-scaling factor $\lambda$ and setting $\lambda =
1$ afterwards gives rise to the differential equation
\begin{equation}
\label{gl:diffglfuerssing}
    s_{\sts \mb{s}}(\varepsilon,l) = a_{\varepsilon}
    \varepsilon \partial_{\varepsilon}s_{\sts \mb{s}}(\varepsilon,l) +
    \frac{1}{d}\,l\partial_ls_{\sts \mb{s}}(\varepsilon,l)
\end{equation}
for the singular part of the reduced entropy.

The entropy of a finite system is analytic \cite{gros04} and therefore 
$s_{\sts \mb{s}}$ can be expanded with respect to
the energy deviation $\varepsilon$, yielding the series
expansion
\begin{equation}
\label{eq:Mffs_entwickelt}
    s_{\sts \mb{s}}(\varepsilon,l) = B_0(l) + \sum_{n=1}^{\infty}
    \frac{1}{n}B_n(l)\varepsilon^{n}.
\end{equation}
{From} equation
(\ref{gl:diffglfuerssing}) the differential equation
\begin{equation}
\label{gl:zuvergleichen}
    B_n(l) = na_{\varepsilon} B_n(l) + \frac{1}{d}\, l \partial_l
    B_n(l)
\end{equation}
is obtained for the expansion coefficients $B_n(l)$, $n=0,1,\ldots$
This differential equation
has the solution
\begin{equation}
    B_n(l) = B_n^{(0)}l^{d(1-na_{\varepsilon})}
\end{equation}
where the $B_n^{(0)}$ are size-independent coefficients. Similarly, the regular
part of the reduced entropy
can be expanded into the
series
\begin{equation}
\label{eq:regsred_entwickelt}
    s_{\sts \mb{r}}(\varepsilon,l) = C_0(l) + \sum_{n=1}^{\infty}
    \frac{1}{n}C_n(l)\varepsilon^{n}.
\end{equation}
As the regular part of the entropy of the infinite system is also
analytic, it is natural to assume that the coefficients are
regular functions in $l$ so that the $C_n(0)$ are the expansion
coefficients of the regular part of the entropy of the infinite
system. Mathematically speaking, the limiting procedure $l\to 0$
and the summation in (\ref{eq:regsred_entwickelt}) can be
interchanged. This assumption is not possible for the singular
part whose limit in the infinite system is non-analytic. 

Taking everything together we obtain that the
expansion of the total entropy of a finite system is of the form
\begin{equation}
\label{eq:ganzesred_entwickelt}
    s(\varepsilon,l) = s_0(l) + \sum_{n=1}^{\infty}
    \frac{1}{n}\left(B_n^{(0)}l^{d(1-na_{\varepsilon})} + C_n(l)\right)\varepsilon^{n}.
\end{equation}
To proceed further let us consider the vicinity of the pseudo-critical point 
$e_{\sts \mb{pc}}$ of the
finite system of inverse length $l$. From a Taylor expansion we obtain (with
$\varepsilon= e - e_{\sts \mb{pc}}$)
\begin{equation}
\label{eq:ganzesred_entwickelt_vier}
    s(\varepsilon,l) = s_0(l) + \beta_{\sts \mb{pc}}\varepsilon + \frac{1}{2}A_2(l)\varepsilon^2 +
    \frac{1}{4}A_4(l)\varepsilon^4 +\ldots
\end{equation}
where the coefficients of the second and fourth order terms,
\begin{equation}
    A_2(l) = B_2^{(0)}l^{\frac{\alpha_{\varepsilon}}{\nu_{\varepsilon}}}
    + C_2(l)
\end{equation}
and
\begin{equation}
A_4(l) = B_4^{(0)}l^{\frac{\alpha_{\varepsilon}-2}{\nu_{\varepsilon}}}
    + C_4(l),
\end{equation}
involve the critical exponents of the microcanonical system.

\begin{figure}[h!]
\begin{center}
\epsfig{file=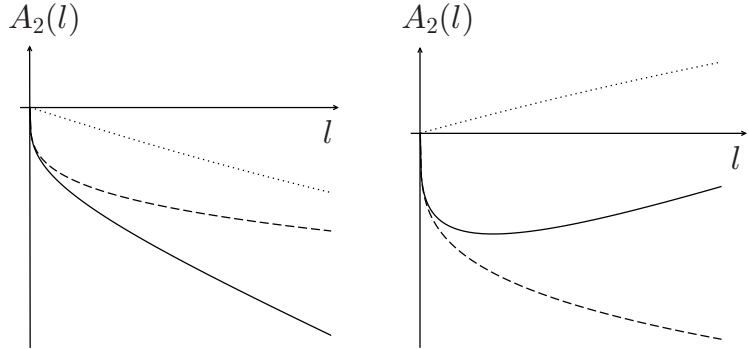,width=13em, angle=270}
\caption[Schematic discussion the possible size-dependence of the
expansion coefficient $A_2(l)$ of the reduced entropy
function]{\label{bild:schema_koeffevolution}\small Schematic
discussion of the possible  size-dependence of the expansion
coefficient $A_2(l)$ (solid lines) of the reduced entropy
function. Shown are two cases, namely a negative (left) and
positive (right) contribution $C_2(l)$. For  small
$l$ the variation is determined by the sum of the singular
contribution $B_2(l)$ governed by the ratio
$\alpha_{\varepsilon}/\nu_{\varepsilon}$ (dashed lines) and the
contributions  from  $C_2(l)$ (dotted lines). For a rapidly
growing positive contribution $C_2(l)$ the coefficient can have a
minimum (right). }
\end{center}
\end{figure}

The coefficient $A_2(l)$ of the second degree term is of
particular interest as it describes the evolution of the
microcanonical specific heat at the
pseudo-critical energy $e_{\sts \mb{pc}}$. Indeed, the curvature at
$e_{\sts \mb{pc}}$ as a function of the system size is given by
\begin{equation}
\label{gl:kruemmungohnekor}
    \partial_{\varepsilon\varepsilon}s(\varepsilon=0,l) =A_2(l)= B_2^{(0)}l^{\frac{\alpha_{\varepsilon}}{\nu_{\varepsilon}}}
    + C_2(l) < 0.
\end{equation}
As the function $C_2(l)$ is regular and has to vanish in the
thermodynamic limit ($l\to 0$) in order to produce a diverging
specific heat at the critical point $e_{{\sts \mb{c}},\infty}$ of
the infinite system, it has to be of the form
\begin{equation}
    C_2(l) = v_1l + \frac{1}{2}v_2l^2 +\ldots
\end{equation}
for small $l$. 
For a continuous phase transition the coefficient $A_2(l)$ is negative for all 
inverse system sizes $l$ (see equations (\ref{gl:kruemmungohnekor}) and 
(\ref{eq:spezallgfuermikro})), therefore, 
the coefficient $B_2^{(0)}$ is also negative as it is
the dominating one for the asymptotic limit of vanishing $l$.
However, the sign of the coefficient $C_2(l)$ is not further
restricted. The possible evolutions of the coefficient $A_2(l)$
depending on the sign and variation of $C_2(l)$ are schematically
shown in figure \ref{bild:schema_koeffevolution}.
A back-bending of the function $A_2(l)$ for decreasing system
sizes can be caused by a large enough positive coefficient
$C_2(l)$. The resulting minimum in $A_2$(l)
has the consequence that the maximum of the specific
heat of small systems decreases with increasing systems size $L$
and increases again in the limit of large systems. This is exactly what we observe numerically.
Thus, the phenomenological
viewpoint developed in this section accounts for the peculiar behaviour of the
specific heat
of small microcanonical systems reported in section
\ref{kap:untersuchespezskal}. 
 
\begin{figure}[h!]
\begin{center}
\epsfig{file=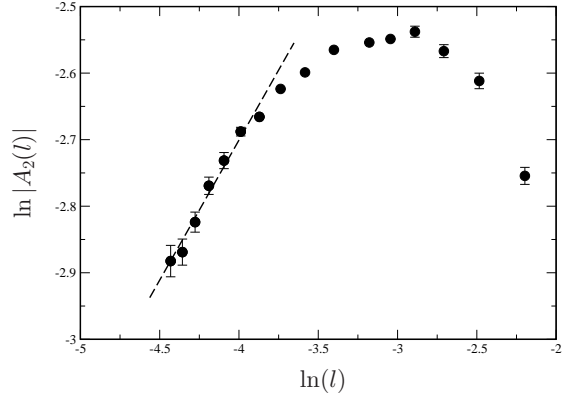,width=13em, angle=270}
\caption[Double-logarithmic plot of the coefficient $|A_2(l)|$ of
the three-state Potts model in two dimensions with periodic
boundary conditions ]{\label{bild:po2d_log_koeffevolution}\small
 Double-logarithmic plot of the coefficient $|A_2(l)|$ of
the three-state Potts model in two dimensions with periodic
boundary conditions. For large systems the data is approximately
given by a straight line. The linear regression for the data of
the largest systems ($L$ between 48 and 84) gives the slope
$0.42\pm 0.03$ (dashed line).}
\end{center}
\end{figure}

Finally, let us note that the
consideration of the limit $l\to 0$ (i.\,e. $L\to\infty$) allows
in principle the determination of the ratio
$\alpha_{\varepsilon}/\nu_{\varepsilon}$ (see \cite{hove04} for another
recent discussion of this point).
In the limit  of large systems the evolution of
the coefficient $A_2(l)$ as a function of the inverse system size
$l$ is governed by the ratio
$\alpha_{\varepsilon}/\nu_{\varepsilon}$. In figure
\ref{bild:po2d_log_koeffevolution} a double-logarithmic plot of
the coefficient $|A_2(l)|$ is shown for the Potts system with
periodic boundary conditions.
The data for large systems is in good approximation described by a
straight line. The slope of this line is an estimate of the
exponent ratio $\alpha_{\varepsilon}/\nu_{\varepsilon}$. From our
data we obtain the slope $0.42\pm 0.03$ which
has to be compared with the exactly known value $2/5$. The
evolution of the coefficient $A_2(l)$ for large system sizes is
indeed determined by the critical exponent ratio
$\alpha_{\varepsilon}/\nu_{\varepsilon}$. Hence, the
microcanonical analysis of the evolution of physical quantities of
finite systems allows, in principle, the determination of the true
critical exponents characterising the critical behaviour of the
infinite system (see \cite{huel02,plei04} for a discussion of how to
determine the order parameter critical exponent directly from the 
density of states of small systems). 
 
The picture is somehow different for the systems with open boundary
conditions. There, the asymptotic regime governed by the exponent
$\alpha_{\varepsilon}/\nu_{\varepsilon}$ is not yet reached for
the considered systems sizes ($L \leq 84$). This probably has its
origin in the finite-size contributions of the free surfaces \cite{plei04b}
which strongly affect
the behaviour of small systems with
open boundaries.
In order to  determine the ratio
$\alpha_{\varepsilon}/\nu_{\varepsilon}$ in this case
much larger system sizes have to be investigated.

The phenomenological picture developed so far applies only to systems
with an algebraically diverging specific heat in the thermodynamic limit.
For systems with a logarithmically  diverging specific heat (as encountered
for example in the two-dimensional Ising model), the
canonical and microcanonical critical exponents are identical.
This suggests that the finite-size behaviour of the various
physical quantities at the pseudo-critical point should be described by
the same asymptotic law in terms of the system size in both
ensembles. As the microcanonical specific heat at $e_{\sts \mb{pc}}$
is basically given by the inverse of the coefficient $A_2(l)$, the
asymptotic behaviour  of the
canonical specific heat~\cite{ferd69} suggests the form
\begin{equation}
\label{gl:a2koeff_fuerlog}
    1/A_2(l) = \tilde{B}_2^{(0)}\ln L + \tilde{C}_2(l).
\end{equation}
Plotting 
$1/A_2(l)$ against $\ln L$ as done in Figure \ref{bild:is2d_koeffevolution}b for the
two-dimensional Ising model with open boundaries shows that the asymptotic
law (\ref{gl:a2koeff_fuerlog}) indeed holds for large system sizes.

To conclude the phenomenological considerations of this subsection a short remark about the corrections
to scaling due to irrelevant scaling fields must be added. These
additional correction terms are non-analytic in the inverse system
size $l$ and alter therefore the size-dependence of the
coefficient $B_2(l)$. Denoting the non-integer exponent of the
correction to scaling term  by $\omega_{\varepsilon}$, the
expression of $A_2(l)$ for a system with an algebraically
diverging specific heat is given by (compare relation
(\ref{gl:kruemmungohnekor}))
\begin{equation}
    A_2(l) =
    B_2^{(0)}l^{\frac{\alpha_{\varepsilon}}{\nu_{\varepsilon}}}\,(1 + b_1l^{\omega_{\varepsilon}} + \ldots)
    + C_2(l).
\end{equation}
Here $C_2(l)$ is again the correction term that arises from the
regular part of the entropy. Note that a negative coefficient
$b_1$ with suitably large modulus may also cause the possible
back-bending of the coefficient $A_2(l)$ for small system sizes.

\section{Conclusions}

Precursor effects of phase transitions can be very different in the microcanonical
and in the canonical treatment of finite systems. The best known example is the
appearance of a negative microcanonical specific heat in small systems that 
announces a discontinuous phase transition. But typical features are also
encountered in the microcanonical ensemble in cases where a continuous
phase transition takes place in the thermodynamic limit, the most intriguing being 
a divergent susceptibility already present in finite systems.

In this work we have shown that the microcanonical specific heat also shows a peculiar
behaviour for small systems that undergo  a continuous phase transition in the
thermodynamic limit. The observed initial decrease of the specific heat 
for increasing system sizes has to be
compared to the behaviour in the canonical ensemble where a monotonic increase of
the maximum of the specific heat is encountered.

We have presented a phenomenological finite-size scaling theory that permits to explain
this peculiar behaviour. This theory, which is based on the analyticity of the microcanonical
entropy surface, uses as a variable the distance to the pseudo-critical point 
$e_{\sts \mb{pc}}$ of a given finite system. This unusual ansatz has allowed us
recently to extract the order parameter critical exponent directly from the density
of states of small systems \cite{plei04}. The phenomenological finite-size scaling theory
should therefore be viewed in the broader context of deriving a finite-size
scaling theory in the microcanonical ensemble.

There do exist some earlier attempts at a microcanonical finite-size
scaling theory. A
finite-size scaling theory for a microcanonical ensemble with the
energy as its only natural variable was also formulated in
\cite{bruc99}. However, that work is based on a definition
of the microcanonical entropy of finite systems that is different
from the definition (\ref{eq:def_entropie}) used in our work. In fact,
the definition of
the entropy used in \cite{bruc99} has a major disadvantage.
It is well known that 
the various statistical ensembles can be formulated in a unified
way in terms of the extremal properties of Boltzmann's
eta-function. These extremal properties have to be worked out
under certain subsidiary conditions which are related to the way
how the system is coupled to its environment in the different
ensembles. This unified point of view is, however, not
possible for the microcanonical ensemble considered in
\cite{bruc99}. 

Microcanonical finite-size scaling relations
were also considered in \cite{kast00,kast01} for the whole
entropy surface $s(e,m,L^{-1})$. In
those works the analysis of the entropy surface
$s(e,m,L^{-1})$ was carried out with respect to the transition
point of the infinite system. This is different in the
considerations of the present work where the relative deviation from the {\em finite}
system transition point has been  investigated.  Microcanonical
finite-size scaling relations were also investigated in \cite{desa88}.
In that work the microcanonical quantities
were basically defined as expectation values with respect to the
microcanonical probability distribution
$p_E(M) \sim \Omega(E,M)$. The
microcanonical quantities analysed in the present work are defined
in a conceptionally different manner.

Finally, let us note that in experiments on nuclear systems or atomic clusters
knowledge of the infinite system is usually not available. Therefore, our
scaling theory involving only quantities of the finite system considered 
seems to be the most appropriate for describing this kind of experiments.

\end{document}